\documentstyle[12pt]{article}
\def\beq{\begin{equation}}
\def\eeq{\end{equation}}
\def\bea{\begin{eqnarray}}
\def\eea{\end{eqnarray}}
\def\bq{\begin{quote}}
\def\eq{\end{quote}}

\def\cpsub{\setlength{\unitlength}{8pt}\begin{picture}(2,1)\mbox{\scriptsize CP}\put(-1.6,-0.1){\line(2,1){1.7}}\end{picture}}

\def\cp1sub{\setlength{\unitlength}{8pt}\begin{picture}(2,1)\mbox{\scriptsize CP} \end{picture}}

\def\gappeq{\mathrel{\rlap {\raise.5ex\hbox{$>$}}
{\lower.5ex\hbox{$\sim$}}}}

\def\lappeq{\mathrel{\rlap{\raise.5ex\hbox{$<$}}
{\lower.5ex\hbox{$\sim$}}}}

\begin{document}
\topmargin -0.5cm
\oddsidemargin -0.3cm
\pagestyle{empty}
\vspace*{5mm}
\begin{center} {\bf THE CP-CONSERVING DIRECTION}\\   
\vspace*{1cm}  {\bf M.C. Ba\~{n}uls} \footnote{IFIC, Centro
Mixto Univ. Valencia - CSIC, E-46100 Burjasot (Valencia), Spain;
banuls@titan.ific.uv.es}
and  {\bf  J. Bernab\'eu} 
\footnote{Departamento de F\'{\i}sica Te\'orica, Univ. Valencia, E-46100
Burjassot (Valencia), Spain; bernabeu@evalvx.ific.uv.es} \\
\vspace*{1cm}
\vspace*{2cm}   {\bf ABSTRACT} \\ \end{center}
\vspace*{5mm}
\noindent 
A symmetry transformation is well defined in the case of an invariant theory, being the corresponding operator undetermined otherwise.
However, we show that, even with CP violation, it is possible to determine the CP transformation by separating the Lagrangian of the Standard Model in a CP-conserving and a CP-violating part, in a unique way, making use of the empirically known quark mixing hierarchy. 
To $O(\lambda^3)$ for the $B_d$-system, the CP conserving direction matches one of the sides of the $(bd)$ unitarity triangle.
We use this determination to calculate the rephasing invariant parameter $\varepsilon$, which measures CP-mixing in the $B^0-\bar{B}^0$ system.

\vspace*{5cm}

\vfill\eject

\setcounter{page}{1}
\setcounter{footnote}{0}
\pagestyle{plain}

\section{Introduction}

The invariance of a theory under a certain symmetry operation means that there exists a way to choose the transformation of fields such that the corresponding action, $A= \int d^4 x L(x)$,  remains unchanged.
Thus, the requirement that the Lagrangian commutes with the symmetry operator determines the transformation of the fields.
In particular, in a CP invariant theory, the CP transformation is well defined by the symmetry properties of the corresponding Lagrangian.

Even if the theory is not invariant, there are still cases in which such a determination is also possible.
This is the situation when the structure of the Lagrangian allows the unambiguous separation of a CP conserving part which includes flavour mixing, from a different interaction, responsible for the CP non-invariance.
In this type of models, as in superweak interaction \cite{SW}, the invariant part of the Lagrangian determines the action of the symmetry operation on the fields.

The case we are interested in is that of the Standard Model: a theory where flavour mixing and CP violation cannot be separated in that way.
Therefore there is no phase choice for the transformed fields which leaves the Lagrangian invariant, and different choices of phases, and thus of CP operation, will yield different observables.
We are going to show, however, that in this CP violating scenario the theory admits a well defined CP transformation by separating the Lagrangian in a CP-conserving and a CP-violating part in a unique way, making use of the empirically known quark mixing hierarchy.
In such a description,
the phases introduced by the CP transformation of physical fields are determined, and define the so called CP-conserving direction.
The problem of CP violation admits then a geometrical interpretation in the complex plane of unitarity triangles, with reference to this direction.

In the next section, we revise the definition of CP transformation and study the symmetry properties of the different pieces in the SM Lagrangian.
The concept of CP-conserving direction is introduced in section~\ref{sec:CPdir}, associated to the unitarity triangles.
In section~\ref{sec:det} we show how the CP transformation is determined by the theory itself, due to the quark mixing hierarchy.
Finally, in section~\ref{sec:apli} we discuss how this unambiguous determination allows the performance of a CP-tag which gives access to the experimental measurement of the CP-mixing parameter $\varepsilon$.

\section{The CP transformation}

The CP symmetry operation is defined for free fields, by requiring that the transformed and the original fields satisfy the same equations of motion. 
This does not determine the action of CP operation completely, since an arbitrary phase can be included in the definition of each transformed field with no effect on the equations.

When the theory contains N families of fermions with the same flavour charges, the generalized CP operation involves, instead of these arbitrary phases, a unitary $N \times N$ matrix $\Phi$, acting on flavour space.
\bea
\psi (\vec{x},t) & \stackrel{\rm CP}{\rightarrow} & \Phi \gamma_0 {\cal C} \bar{\psi}^T(-\vec{x},t) \nonumber \\
\bar{\psi}(\vec{x},t)& \stackrel{\rm CP}{\rightarrow} & -\psi^T(-\vec{x},t) {\cal C}^{-1} \gamma_0 \Phi^+ \ , \label{Eq.1}
\eea
where $\cal C$ is a unitary matrix satisfying the condition ${\cal C}^{-1} \gamma_{\mu} {\cal C} = - \gamma_{\mu}^T$ and usually realized by ${\cal C} = i \gamma^2 \gamma^0$.
Equation (\ref{Eq.1})
is understood for the up-sector, and there is a similar transformation law for the down-sector, involving an independent matrix $\Phi'$.
The invariance condition for strong and electromagnetic interactions requires these matrices $\Phi$, $\Phi'$ to be unitary, leaving them otherwise unfixed.

The Lagrangian is said to be CP-invariant 
if it is possible to find 
a pair of matrices $\Phi$ and $\Phi'$, such that the action does not change under the above transformation.

The Lagrangian density of the Electroweak Standard Model can be symbollically written as \cite{Jar89}:
\beq
L=L(f,G)+L(f,H)+L(G,H)+L(G)-V(H)
\eeq
where $f$ represents the fermions, $G$ the gauge bosons and $H$ the scalar doublet in the theory. 

The first term on the r.h.s., $L(f, G)$, which represents fermion gauge interactions, explicitly breaks $P$, since left-handed and right-handed fields do not interact in the same way.
It also violates $C$, for it involves both vector and axial couplings. 
But it can be shown to be CP-invariant~\cite{GR97}, since a transformation like~(\ref{Eq.1}), with $\Phi=\Phi'$ leaves the Lagrangian invariant. 
This means that up and down-type quarks must transform in identical way.
$L(G,H)$ is CP-conserving, too, since the phases in the scalar field transformation can be properly chosen. 
Thus without the scalar sector, the theory would be CP-invariant.

However, when we consider the whole Lagrangian, we also need to study the transformation properties of $L(f,H)$, the interaction of fermions with the scalar doublet.
After spontaneous symmetry breaking what remains for this term is:
\beq
L(f,H) \rightarrow -\sum_{j,k=1}^{N} \left \{m_{jk}\bar{q_j}_L {q_k}_R+m'_{jk}{\bar{q_j}'}_L {q_k}'_R + h.c. \right \} \left ( 1+\frac{H}{v}\right ) \ ,
\label{SSB}
\eeq
where $N$ is the number of families, $q$ and $q'$ are the N-component vectors representing weak eigenstates for up and down quarks, and $m_{jk}$, $m_{jk}'$ are the matrix elements of $M$ and $M'$, the complex quark mass matrices.
These matrices are obtained from Yukawa couplings after SSB, and need to be neither real nor hermitian. 
However, due to the structure of gauge interactions any unitary transformation on the right-handed quarks is unobservable, and it is possible to restrict ourselves to hermitian mass matrices without loss of generality~\cite{FJ85}. 

So far we are in a weak basis, dealing with non-physical quark fields. Here, the quark mass term and the charged-current interaction term read~\cite{BBG86}:
\beq
L=\left ( \begin{array}{c c} \bar{q} & \bar{q'} \end{array} \right )_L \left ( \begin{array}{c c} M & 0 \\ 0 & M' \end{array} \right ) \left ( \begin{array}{c} q \\ q' \end{array} \right )_R + g{\bar{q}}_L \gamma_{\mu} {q'}_L W^{\mu}+h.c.
\label{weak.basis}
\eeq

We want to study the properties of the Lagrangian (\ref{weak.basis}) under a general CP transformation for up- and down-type quarks. 
Since the transformation is defined up to the unitary matrices $\Phi$, $\Phi'$,
the question is whether there is a way of choosing them leaving $L$ invariant.
Imposing CP-invariance to the mass term, we get the following condition on the CP matrices:
\bea
M^*&=&\Phi^+ M \Phi \nonumber \\
{M'}^*&=&{\Phi'}^+ M' \Phi' \ .
\label{mass.inv}
\eea

Being hermitian, $M$ and $M'$ can be diagonalized by unitary matrices:
\bea
M &=& U^+ D \, U \nonumber \\
M' &=& {U'}^+ D' U' \ ,
\label{diagon}
\eea
with $D={\rm diag} (m_u, m_c, m_t)$ and $D'={\rm diag} (m_d, m_s, m_b)$.
The physical quark fields are thus given by:
\beq
q_{phys} \equiv U q \  \hskip2cm q_{phys}' \equiv U' q' \ .
\label{q.phys}
\eeq

It is then immediate to find matrices $\Phi$, $\Phi'$ which satisfy the condition (\ref{mass.inv}), namely:
\beq
\Phi=U^+ e^{2 i \Theta} U^* \  \hskip1.5cm 
\Phi'={U'}^+ e^{2 i \Theta^{'}} {U'}^* \ ,
\label{CP.weak}
\eeq
where $\Theta$ and $\Theta^{'}$ are real diagonal matrices.

We may now perform the transformation (\ref{CP.weak}) on the Lagrangian (\ref{weak.basis}). 
The mass term is invariant by construction of $\Phi$, $\Phi'$. 
However, for the charged current term we get:
\beq
{\bar{q_i}}_L \gamma_{\mu} {q_j'}_L \stackrel{\rm CP}{\rightarrow} -{\bar{q_j'}}_L {\Phi'}_{jk}^T \Phi_{ki}^*\gamma^{\mu} {q_i}_L 
\eeq

We can define a unitary matrix:
\beq
B \equiv \Phi {\Phi'}^+ 
\label{B.def}
\eeq
and write the transformed charged current term as:
\beq
(C \! P) L_{CC} (C \! P)^{-1} = \sum_{i,j} g B_{ij}{\bar{q_i}}_L \gamma_{\mu} {q_j'}_L W^{\mu}+h.c. \label{Eq.11}
\eeq

The only way for the Lagrangian (\ref{weak.basis}) to be invariant is to have $B=I$, and any difference will measure CP violation.

It is always possible \cite{tesina} to find a weak basis, without any change in physics, where $B$ is diagonal: \mbox{$B ={\rm diag} (b_1,b_2,b_3)$}.
Its eigenvalues will satisfy $|b_i|=1$ \cite{Gan66}. 
This suggests the interpretation of $B$ as the relative phase between CP transformations of weak up and down quarks, as defined by mass term invariance (\ref{mass.inv}).
In absence of charged currents, this relative phase would be unobservable, but with the charged current interaction, it gives rise to CP violation, according to (\ref{Eq.11}).

Therefore we observe:
\begin{enumerate}
\item{We may always choose a CP transformation, i.e. a pair of unitary $\Phi$, $\Phi'$ as given by (\ref{CP.weak}) such that the weak mass term (plus the strong and electromagnetic interactions) will be CP conserving.
This definition, however, does not fix completely  the action of CP on the quark fields: the phases $\Theta$ and $\Theta^{'}$ are arbitrary and not defined by the invariance requirement.
In absence of the charged current term, this arbitrariness would not involve any ambiguity in the definition of CP, since no interaction would be sensitive to the relative phase between up and down sectors.}
\item{If the mass matrices were such that one could choose this transformation with $\Phi=\Phi'$, the charged-current term would be invariant, too, since $B=1$. 
Otherwise, different CP transformations for up and down sectors in the weak basis result in CP violation.}
\item{CP invariance in a non-abelian gauge theory fixes the CP phases.
If CP is advocated as coming from a superweak interaction, then the CP operation is completely fixed by the Standard Model Lagrangian (\ref{weak.basis}), which respects in that case the symmetry.
The necessary and sufficient condition for CP invariance in the Standard Model \cite{Jar85} leads to the phase fixing \cite{BBG86}.
But, in general, for a non-invariant Lagrangian (\ref{weak.basis}) the phases $\Theta$ and $\Theta^{'}$ remain unfixed.
Any choice of them would leave the weak mass term invariant, but the CP-violating quantities we could construct would depend on how we had defined the operation.
In the next sections, we show how it is possible to separate also the CP conserving part of the charged current term in a non-ambiguous way, as far as we work with the Lagrangian at order ${\cal O}(\lambda^3)$. 
This eliminates the arbitrariness in the definition of the CP transformation.}
\end{enumerate}

\section{The CP-conserving direction}
\label{sec:CPdir}

In the physical basis given by (\ref{q.phys}):
\bea
\Phi & \rightarrow & U \Phi \, U^T=e^{2 i \Theta} \nonumber \\ 
\Phi' & \rightarrow & U' \Phi' {U'}^T=e^{2 i \Theta^{'}}
\label{CP.phys} 
\eea

Fixing the intermediate phases in the transformation of weak quarks is then equivalent to the choice of the (diagonal) CP transformation of physical fields.
In this basis, (\ref{weak.basis}) reads:
\beq
L=\left ( \begin{array}{c c} \overline{q}_{phys} & \overline{q}_{phys}' \end{array} \right ) \left ( \begin{array}{c c} D & 0 \\ 0 & D' \end{array} \right ) \left ( \begin{array}{c} q_{phys} \\ q_{phys}' \end{array} \right )+ g {{\overline{q}}_{phys}}_L \gamma_{\mu}V {q_{phys}'}_L {W^+}^{\mu}
\eeq
with $V \equiv U {U'}^+$, the quark mixing matrix.
The only term which is not trivially invariant under CP for the CP-phases (\ref{CP.phys}) is the charged current interaction, which transforms according to:
\beq
V \stackrel{\rm CP}{\rightarrow} e^{2 i \Theta} V^* e^{-2 i \Theta^{'}}
\eeq
Therefore, by means of a change of basis, we have moved all the CP problem to the charged current term.

Once the CP operation has been fixed by a certain choice of CP-phases, we can separate the Lagrangian in a unique way into a CP-even and a CP-odd part:
\bea
L_{\cp1sub} \equiv \frac{1}{2} \left ( L + (CP) L (CP)^{-1} \right )
\nonumber \\
L_{\cpsub} \equiv \frac{1}{2} \left ( L - (CP) L (CP)^{-1} \right )
\label{CP.part.L}
\eea

Similarly, we can separate the mixing matrix in two parts:
\beq
V=V_{\cp1sub}+V_{\cpsub} \label{Eq.16}
\eeq
each of them corresponding to the coefficients of charged-current term in $L_{\cp1sub}$ and $L_{\cpsub}$, respectively.

If the theory is CP invariant, the decomposition is trivial, since $L_{\cp1sub}\equiv L$. In any other case, we obtain different separations depending on how we define the symmetry operator CP.
For the transformation (\ref{CP.phys}), we get:
\bea
V_{\cp1sub} \equiv\frac{1}{2} \left (V+e^{2 i \Theta} V^* e^{-2 i \Theta^{'}} \right )= \frac{1}{2} U \left ( 1 + B \right ) {U'}^+
\nonumber \\
V_{\cpsub} \equiv \frac{1}{2} \left (V-e^{2 i \Theta} V^* e^{-2 i \Theta^{'}} \right )= \frac{1}{2} U \left ( 1 - B \right ) {U'}^+
\label{CP.part.V}
\eea
which shows again the role of the unitary matrix $B$ for CP-violation.
Whereas $V \neq I$ describes flavour mixing, $B \neq I$ describes CP violation.

In spite of $V$ being a unitary matrix, its CP-even and CP-odd projections are not. On the contrary, one can show that $V_{\cp1sub}$ is unitary if and only if $V_{\cpsub}=0$, i.e.~if there is no CP violation.

This suggests one possible way to quantify CP violation by calculating the norm of $V_{\cpsub}$. 
Any non-zero value of $\|V_{\cpsub} \|$ will be a signal of CP violation, independent of the quark basis.
We define:
\beq
\| V_{\cpsub} \| = \limsup_{\| \Psi \|=1 }  \| V_{\cpsub} \Psi \|
\eeq

Due to the unitarity of $U$, $U'$:
\beq
\| V_{\cpsub} \| = \frac{1}{2} \| 1 - B \|
\eeq

If we call \mbox{${\rm Re}(b) \equiv {\rm min} \{ {\rm Re} (b_i)\}$}, we get:
\beq
\|V_{\cpsub}\|^2 = \frac{1}{2} \left(1-{\rm Re} (b) \right)
\eeq

Therefore, $\| V_{\cpsub} \|$ will be null if and only if $B\equiv I$, i.e.~if CP is conserved.

However, as we have seen, and due to the non-invariance of the theory, the definition of CP is not unique. 
We could give different prescriptions resulting in different sizes of ``CP violation'', as this would measure the non-invariance of $L$ under different transformations, associated with different CP-phases.

The definition of the CP-conserving direction provides a more practical way of studying the CP violation.
We may write (\ref{CP.part.V}) as:
\bea
V_{\cp1sub}=e^{i \Theta} {\rm Re} \left (e^{-i \Theta} V e^{i \Theta^{'}} \right ) e^{-i \Theta^{'}}
\nonumber \\
V_{\cpsub}=i e^{i \Theta} {\rm Im} \left (e^{-i \Theta} V e^{i \Theta^{'}} \right ) e^{-i \Theta^{'}}
\eea

Representing $V$ matrix elements as vectors in the complex plane, we notice $(V_{\cp1sub})_{ij}$ and $(V_{\cpsub})_{ij}$ are the projections of $V_{ij}$ on the direction of $e^{i(\theta_i - \theta_j^{'})}$ and its orthogonal $i e^{i(\theta_i - \theta_j')}$, which are fixed by the operator choice.

In the Standard Model, with three families, the quark mixing matrix $V$ is a unitary $3 \times 3$ matrix described by four independent parameters. 
The unitarity condition $V^+V=V\, V^+=I$ yields six off-diagonal relations which can be represented in the complex plane by six triangles. 
Under rephasing of the quark fields, these triangles change their orientation in the plane, but their shape is invariant. 
Therefore, the triangles are physical objects, whose angles and sides can be measured.

From the condition $V^+ V=I$ we obtain three triangles for the down sector:
\beq
\sum_{k=1}^{3}V_{ki}^* V_{kj} = 0, \hskip2cm ij=ds, \, sb, \, db
\eeq

Each side (fixed $k$) of one of these triangles can be expressed as:
\beq
V_{ki}^* V_{kj} = e^{i(\theta_i' - \theta_j')} \left \{{\rm Re} \left (e^{-i \theta_i'} V_{ki}^* V_{kj} e^{i \theta_j'} \right )+i{\rm Im} \left (e^{-i \theta_i'} V_{ki}^* V_{kj} e^{i \theta_j'} \right ) \right \}
\label{V*V.proj}
\eeq

If we use $V$ decomposition (\ref{Eq.16}) into CP-conserving and CP-violating parts, we may write
\beq
V_{ki}^* V_{kj} = \left [(V_{\cp1sub})_{ki}^*(V_{\cp1sub})_{kj}+(V_{\cpsub})_{ki}^*(V_{\cpsub})_{kj}\right ]+\left [(V_{\cp1sub})_{ki}^*(V_{\cpsub})_{kj}+(V_{\cpsub})_{ki}^*(V_{\cp1sub})_{kj}\right ]
\eeq

Then the first term in (\ref{V*V.proj}) can be read as the CP-conserving part of the triangle side, and corresponds to the projection of $V_{ki}^* V_{kj}$ on the direction defined by $e^{i(\theta_i'-\theta_j')}$.
We thus call this $e^{i(\theta_i'-\theta_j')}$ the {\itshape CP-conserving direction} associated to the $(ij)$ triangle.

Therefore, every one of the six triangles has an associated CP direction. 
However, not all of them are independent, due to the cyclic relations \footnote{The flavour subindex already specifies the quark sector}:
\bea
e^{i(\theta_b-\theta_d)}=e^{i(\theta_b-\theta_s)}e^{i(\theta_s-\theta_d)} \nonumber \\
e^{i(\theta_t-\theta_u)}=e^{i(\theta_t-\theta_c)}e^{i(\theta_c-\theta_u)}
\label{phase.const}
\eea

These directions are attached to the triangles, so they would rotate with them under quark rephasing. 
They are not physical by themselves, but the relative phases between triangle sides and them are rephasing invariant, and physical once the CP operator has been fixed.
Since only the relative phases between equally charged quarks will appear in the effective Hamiltonian for neutral mesons, we only need to fix four phases (two for each sector) in order to have a well-defined CP transformation.

\section{Determination of the CP operator}
\label{sec:det}

We may ask whether the theory itself filters a well-defined CP operator. 
In the $K$-system, the CP symmetry is only slightly violated and its size \cite{CC64} is of the order $O(10^{-3})$.
This is understood in the Standard Model as a consequence of the need to involve the three families to generate CP-violation so that its effective coupling contains higher powers of the quark mixing $\lambda$ than that of the CP-conserving flavour mixing $K^0-\bar{K}^0$.
This suggests the idea of searching for a ``natural'' CP definition in the Standard Model, based on the empirically known quark mixing hierarchy.

According to this experimental fact, the magnitude of $V$ matrix elements can be written in terms of a perturbative parameter $\lambda \simeq 0.22$ as:
\beq
V \simeq \left ( \begin{array}{c c c}
1 & \lambda & \lambda^3 \\
\lambda & 1 & \lambda^2 \\
\lambda^3 & \lambda^2 & 1
\end{array}
\right ) \ .
\label{mix.hier}
\eeq

Therefore we can easily estimate the (physical) relative size of every side in the six triangles.
In the down sector:
\bea
\begin{array}[t]{c} V_{ud} V_{us}^* \\ {\scriptstyle O(\lambda)} \end{array} + \begin{array}[t]{c} V_{cd} V_{cs}^* \\ {\scriptstyle O(\lambda)} \end{array} + \begin{array}[t]{c} V_{td} V_{ts}^* \\ {\scriptstyle O(\lambda^5)} \end{array} &=&0 \nonumber \\
\begin{array}[t]{c} V_{us} V_{ub}^* \\ {\scriptstyle O(\lambda^4)} \end{array} +\begin{array}[t]{c} V_{cs} V_{cb}^* \\ {\scriptstyle O(\lambda^2)} \end{array} + \begin{array}[t]{c} V_{ts} V_{tb}^* \\ {\scriptstyle O(\lambda^2)} \end{array} &=&0 \nonumber \\
\begin{array}[t]{c} V_{ud} V_{ub}^* \\ {\scriptstyle O(\lambda^3)} \end{array} +\begin{array}[t]{c} V_{cd} V_{cb}^* \\ {\scriptstyle O(\lambda^3)} \end{array} + \begin{array}[t]{c} V_{td} V_{tb}^* \\ {\scriptstyle O(\lambda^3)} \end{array} &=&0 \ .
\eea

Since we only make use of the empirical fact of hierarchy, this is an experimental result independent of the specific parametrization of CKM matrix, which is rephasing variant.

The first two triangles are much flatter than the third one. 
To order $\lambda^3$, those two triangles collapse to two parallel lines, thus giving CP conservation and a natural fixing of the attached CP directions:
\beq
e^{i(\theta_s-\theta_d)} \equiv \left. \frac{V_{cd} V_{cs}^*}{|V_{cd} V_{cs}^*|} \right |_{O(\lambda^3)} \ , \hskip2cm 
e^{i(\theta_b-\theta_s)} \equiv \left. \frac{V_{cs} V_{cb}^*}{|V_{cs} V_{cb}^*|}\right |_{O(\lambda^3)} \ .
\label{CP.def}
\eeq

Thus the CP invariance requirement on the effective hamiltonian has fixed the corresponding CP-transformation phases of these sectors.
In other words, the choice (\ref{CP.def}) corresponds to CP invariance of $H_{ef\!f}$ for $(sd)$ and $(bs)$ systems to order $O(\lambda^3)$.

One can proceed in a similar way with the up-sector triangles, fixing the CP transformation of up-type quarks.
This procedure completely defines the action of CP operator on the fields, up to a global phase on each sector.

Due to (\ref{phase.const}) the CP conserving direction for the $B_d$ system is already fixed.
This is particularly attractive because the $(bd)$-system keeps a CP-violating triangle to order $\lambda^3$. 
It is of high interest to see the relative orientation between the CP direction and the corresponding unitarity triangle.
\beq
e^{i(\theta_b -\theta_d)}=\left. \frac{V_{cd} V_{cb}^*}{|V_{cd} V_{cb}^*|}\right |_{O(\lambda^3)} \ .
\label{eq:cpdir}
\eeq

Thus, with this CP-phase fixing, the CP conserving direction matches one of the sides of the $(bd)$ triangle to $O(\lambda^3)$.

So far we have not used any particular parametrization, so the resulting CP-direction (or properly its relative direction with respect to the triangle) is physical.
However, it is interesting to study the particular result obtained in the Wolfenstein parametrization \cite{Wol83}, which explicitly describes the experimentally shown hierarchy of mixing:
\beq
V=\left (
\begin{array}{c c c}
1-\frac{\lambda^2}{2} & \lambda & A \lambda^3 (\rho - i \eta)\\
-\lambda & 1-\frac{\lambda^2}{2} & A \lambda^2 \\
A \lambda^3 (1- \rho - i \eta) & - A \lambda^2 &1
\end{array}
\right ) \ .
\label{eq:Wolf}
\eeq

The calculation of~\ref{eq:cpdir} for the representation~\ref{eq:Wolf} of the CKM matrix shows that this parametrization corresponds to the choice of a reference system for the unitarity triangles in which the real axis is placed on the CP-conserving direction. 

The previous phase fixing is unique as long as we neglect CP violation in the $K$ and $B_s$ systems.
To ilustrate this point, we calculate the relevant matrices $B$ and $V_{\cpsub}$ to leading order in $\lambda$. We get, after diagonalization:
\beq
B= \left ( \begin{array}{c c c} 1 & 0 & 0 \\ 0 & 1+ i 2 A \lambda^3 \eta & 0 \\ 0 & 0 & 1- i 2 A \lambda^3 \eta \end{array} \right ) + O(\lambda^4) \ ,
\eeq
and:
\beq
V_{\cpsub}= \left ( \begin{array}{c c c} 0 & 0 &-i A \lambda^3 \eta \\ 0 & 0 & 0 \\-i A \lambda^3 \eta & 0 & 0 \end{array} \right ) + O(\lambda^4) \ ,
\hskip1.5cm  \|V_{\cpsub}\| \simeq A \lambda^3 \eta \ .
\eeq

It is important to notice here that the only perturbative argument makes reference to the quark mixing $\lambda$, and  not to the size of CP violation.
The lowest order, $L_0$, is the Lagrangian to order ${\cal O}(\lambda^3)$ included.
$L_0$ is not CP invariant but the CP-conserving and CP-violating pieces separate in unambiguous manner, thus yielding a well-defined symmetry transformation.
This is enough to describe, to leading order in $\lambda$, CP violation in the $B_d$-system. 
Further precision will be achieved by keeping terms of higher order in $\lambda$ in our Lagrangian.
These new terms would allow us to calculate CP-violating results in the other systems $K$ and $B_s$, without changing the definition of CP operation, since this is given by the well defined CP-conserving part of $L_0$.
In this sense, we may say that CP violation is a perturbative effect in the kaon and $B_s$-systems, because it only appears if higher powers of the expansion parameter $\lambda$ are included.
For the $B_d$-system, on the contrary, CP violation is non perturbative.

\section{Rephasing invariant CP-mixing}
\label{sec:apli}

We are interested in the study of neutral $B_d$-system. 
In the Standard Model, $B^0-\bar{B}^0$ mixing can be described by an effective hamiltonian $H_{ef\!f}^{\Delta B=2}$ obtained from the box diagrams with exchange of two $W$ bosons and whose dispersive part is dominated by top contribution \cite{Bur90}. 

The off-diagonal terms in the $B^0-\bar{B}^0$ mass matrix, $M_{12}$ and $\Gamma_{12}$, are the hermitian (dispersive) and antihermitian (absorptive) parts of the matrix element of $H_{ef\!f}^{\Delta B=2}$ between conjugate mesons:
\beq
M_{12}-\frac{i}{2}\Gamma_{12}=\frac{1}{2 m_B} \langle B^0 |H_{eff}^{\Delta B=2} |\bar{B}^0 \rangle \ .
\eeq

We introduce a third matrix element
\beq
C \! P_{12} = \langle  B^0 | C \! P | \bar{B}^0 \rangle \ ,
\eeq
which depends on quark phase convention and on the definition of CP operator. 
The relative phase between $C \! P_{12}$ and the hamiltonian matrix element $H_{12}$ is, however, invariant under rephasing of quark fields and meson states. 
In Ref. \cite{BB98} we have constructed the rephasing invariant parameter $\varepsilon$, to describe CP-mixing. It reads:
\beq
\varepsilon =\frac{{\rm Im} (\Gamma_{12} C \! P_{12}^*) + 2 i{\rm Im} (M_{12} C \! P_{12}^*) }{2{\rm Re} (M_{12} C \! P_{12}^*)-i {\rm Re} (\Gamma_{12} C \! P_{12}^*) + \Delta m - \frac{i}{2} \Delta \Gamma } \ .
\eeq
The observability of both ${\rm Re} (\varepsilon)$ and ${\rm Im} (\varepsilon)$ has been proved by means of a proposed experiment which starts from a CP-tag of the $B_d$ states.

For the $B_d$-system $\Delta \Gamma$ is very small.
One may then take the limit $\Delta \Gamma \rightarrow 0$, in which ${\rm Re}(\varepsilon)/(1+|\varepsilon|^2)$ vanishes, and the main role in indirect CP violation is played by 
\beq
\frac{{\rm Im}(\varepsilon)}{1+|\varepsilon|^2} \approx \frac{{\rm Im} (M_{12} C \! P_{12}^*)}{\Delta m} \label{Eq.37} \ .
\eeq

To first order in $\alpha_{QCD}$, the effective hamiltonian has a factorised form involving \sloppy \mbox{$(V_{tb}^* V_{td})^2 (\bar{d}b)_{V-A}(\bar{d}b)_{V-A}$}.
In the vacuum insertion approximation, the two hadronic matrix elements which appear in the $B^0-\bar{B}^0$ transition are related through the CP transformation
\beq
\langle B^0 |(\bar{d} b)_{V-A}|0\rangle=
-e^{2i(\theta_d-\theta_b)} C\!P_{12}^* 
\langle 0 |(\bar{d} b)_{V-A}|\bar{B}^0 \rangle^* \ .
\eeq
This relation is explicitly variant under rephasing of quark fields and meson states.

Introducing the so called ``bag parameter'' $B_B$ which equals $1$ only in the vacuum insertion approximation, the dispersive part of the matrix element $M_{12}$ is proportional to
\beq
\langle B^0 |(V_{tb}^* V_{td})^2(\bar{d} b)_{V-A}(\bar{d} b)_{V-A}|\bar{B}^0 \rangle =- \frac{8}{6} {f_{B}}^2 m_{B} B_{B} (V_{tb}^* V_{td})^2 C\!P_{12} e^{-2 i (\theta_d-\theta_b)} \ .
\eeq
This product is invariant under quark rephasing, since the variance of both factors, $(V_{tb}^* V_{td})$ and $e^{- i (\theta_d-\theta_b)}$, exactly cancels.
The only remaining convention dependence is that of $C\!P_{12}$, so that our $\varepsilon$ is rephasing invariant.

For the CP operator given by (\ref{CP.def}), the relevant phase in (\ref{Eq.37}) reads to order $O(\lambda^3)$:
\beq
{\rm arg} (M_{12} C \! P_{12}^*)=2 \, {\rm arg} (V_{tb}^* V_{td} V_{cb} V_{cd}^*) \ ,
\eeq
written in terms of the product of four CKM matrix elements in the SM, so that it is rephasing invariant and determined.

The previous result is valid for any CKM parametrization. 
In particular we get, when expressed in terms of the Wolfenstein parameters $(\rho, \eta)$, the following result:
\beq
\frac{{\rm Im}(\varepsilon)}{1+|\varepsilon|^2} \approx - \frac{\eta (1-\rho)}{(1-\rho)^2+\eta^2} \label{Eq.39} \ .
\eeq

Using the allowed range of values for $\eta$, $\rho$, as found from recent phenomenological analyses of experimental data \cite{PP95}, we find that the value of this quantity lies between -0.37 and -0.18, thus yielding good expectations for the existence of measurable asymmetries in indirect CP violation.
As anticipated, the asymmetries governed by (\ref{Eq.39}) are not perturbative, since both CP-conserving and CP-violating components come from $L_0$.
The observable is of zero order in~$\lambda$.

\section{Conclusions}

In general, the CP transformation is not well-defined unless the theory is invariant or admits a unique splitting into CP-violating and CP-conserving pieces. The SM Lagrangian fulfills none of both conditions so that the CP transformation would be, in principle, partly undetermined, with the consequent ambiguity in the definition of some CP-violating quantities.
However we show that, due to the quark mixing hierarchy, a unique separation as above described is possible in the Lagrangian to ${\cal O}(\lambda^3)$.
The procedure discussed in this paper determines the CP phases for the quark fields transformation and gives rise to a well-defined CP-conserving direction associated to each unitarity triangle.
The problem of CP violation can be now  geometrically interpreted, with reference to this direction.

We want to emphasize that the determination of the CP operator 
for the $B_d$-system 
introduced in this paper is not a product of the authors' 
choice for CP phases, but a consequence of the theory itself.
The Standard Model with hierarchical mixing fixes the CP operator
and manages to separate in a unique way the CP-conserving and the
CP-violating parts of the effective hamiltonian.

The determination of CP-eigenstates of the $B_d$-system
gives to
the rephasing invariant CP-mixing parameter $\varepsilon$
a unique meaning.
It also tells us how to prepare a CP-tag of the $B_d$ states, needed for experiments such as the one discussed in~\cite{BB98}.

Since CP is conserved to order $O(\lambda^3)$ 
in $(bs)$ and $(sd)$ sectors,
a decay amplitude of $B_d$ which is governed by couplings of these sectors, 
or by the $V_{cd}V_{cb}^*$ side of $(bd)$ triangle, 
will not show any CP violation to order $O(\lambda^3)$.
Such a channel is then free of ``direct'' CP violation.

The procedure to perform a CP tag starting from an entangled state $|B \rangle |\bar{B}\rangle$ is then clear.
If a CP-eigenstate of the decay products for one of these channels
is observed at one side of the detector, the other side meson can be identified to be in a CP-eigenstate at the time the first decay happened.
One example of a final state useful for such a preparation is, among others, $J/\psi K_S$, whose decay amplitude is governed by the CP-conserving $(bs)$ triangle.

Nevertheless, not all CP-eigenstates of decay products 
($\pi^+ \pi^-$, $J/\psi K_S$, $D^+ D^-$, etc) are eligible
to perform such a tag.
Since there are CP-conserving decay amplitudes, the separation 
between direct and indirect CP-violation is well defined in each case.
Only the channels whose decay amplitude lies along the
CP conserving direction are free from direct CP violation,
and allow the quantum mechanical preparation of a $B_d$ CP-eigenstate 
through the described mechanism.

Previous discussion shows the feasibility of a CP-eigenstate preparation in a realistic experimental facility.
This 
opens the way for novel experiments, to be performed in B-factories at $\Upsilon(4 S)$, in which the preparation of the system for CP-odd asymmetries is not necessarily based on a flavour-tag, but also on a CP-tag.

\vspace{1cm}

{\bf Acknowledgements}\\

This paper has the benefit of its discussion with many colleagues.
We are particularly grateful to F. J. Botella, J. Donoghue, H. Leutwyler and A. Santamaria for their relevant questions, which helped us to formulate the right problem. 
One of us (M.C.B.) is indebted to the Spanish Ministry of Education and Culture for her fellowship.
This research was supported by CICYT, Spain, under grant No. AEN-96/1718.

\end{document}